\documentclass[aps,english,amsmath,superscriptaddress]{revtex4}

\usepackage{graphicx}
\usepackage{amssymb}
\usepackage{amsmath}
\usepackage{epsfig}
\usepackage[active]{srcltx}

\begin{document}

\title{Nonlocal mechanism for cluster synchronization in neural circuits}

\author {I. Kanter}
\affiliation{Department of Physics, Bar-Ilan University, 52900 Ramat-Gan, Israel}
\author {E. Kopelowitz}
\affiliation{Department of Physics, Bar-Ilan University, 52900 Ramat-Gan, Israel}
\author {R. Vardi}
\affiliation{Gonda Interdisciplinary Brain Research Center, and the Goodman Faculty of Life Sciences, Bar Ilan University, Ramat-Gan, 52900, Israel}
\author {M. Zigzag}
\affiliation{Department of Physics, Bar-Ilan University, 52900 Ramat-Gan, Israel}
\author {W. Kinzel}
\affiliation{Institute for Theoretical Physics, University of Wuerzburg, Am Hubland, 97074 Wuerzburg, Germany}
\author {M. Abeles}
\affiliation{Gonda Interdisciplinary Brain Research Center, and the Goodman Faculty of Life Sciences, Bar Ilan University, Ramat-Gan, 52900, Israel}
\author {D. Cohen}
\affiliation{Gonda Interdisciplinary Brain Research Center, and the Goodman Faculty of Life Sciences, Bar Ilan University, Ramat-Gan, 52900, Israel}

\begin{abstract}
The interplay between the topology of cortical circuits and synchronized activity modes in distinct cortical areas is a key enigma in neuroscience.  We present a new nonlocal mechanism governing the periodic activity mode: the greatest common divisor (GCD) of network loops. For a stimulus to one node, the network splits into GCD-clusters in which cluster neurons are in zero-lag synchronization.  For complex external stimuli, the number of clusters can be any common divisor. The synchronized mode and the transients to synchronization pinpoint the type of external stimuli. The findings, supported by an information mixing argument and simulations of Hodgkin Huxley population dynamic networks with unidirectional connectivity and synaptic noise, call for reexamining sources of correlated activity in cortex and shorter information processing time scales.
\end{abstract}

\maketitle

\section{Inroduction}
The spiking activity of neurons within a local cortical population is typically correlated \cite{1,2,3,4}. As a result, local cortical signals are robust to noise, which is a prerequisite for reliable signal processing in cortex.  Under special conditions, coherent activity in a local cortical population is an inevitable consequence of shared presynaptic input \cite{5,6,7,8,9}. Nevertheless, the mechanism for the emergence of correlation, synchronization or even nearly zero-lag synchronization (ZLS) among two or more cortical areas which do not share the same input is one of the main enigmas in neuroscience \cite{7,8,9}.
It has been argued that nonlocal synchronization is a marker of binding activities in different cortical areas into one perceptual entity \cite{8,10,11,12}. This prompted the hypothesis that synchronization may hold key information about higher and complex functionalities of the network. To investigate the synchronization of complex neural circuits we studied the activity modes of networks in which the properties of solitary neurons, population dynamics, delays, connectivity and background noise mimic the inter-columnar connectivity of the neocortex.

\begin{figure*}[t]
\begin{center}
\includegraphics[angle=0,width=0.945\textwidth,totalheight=0.315\textwidth]{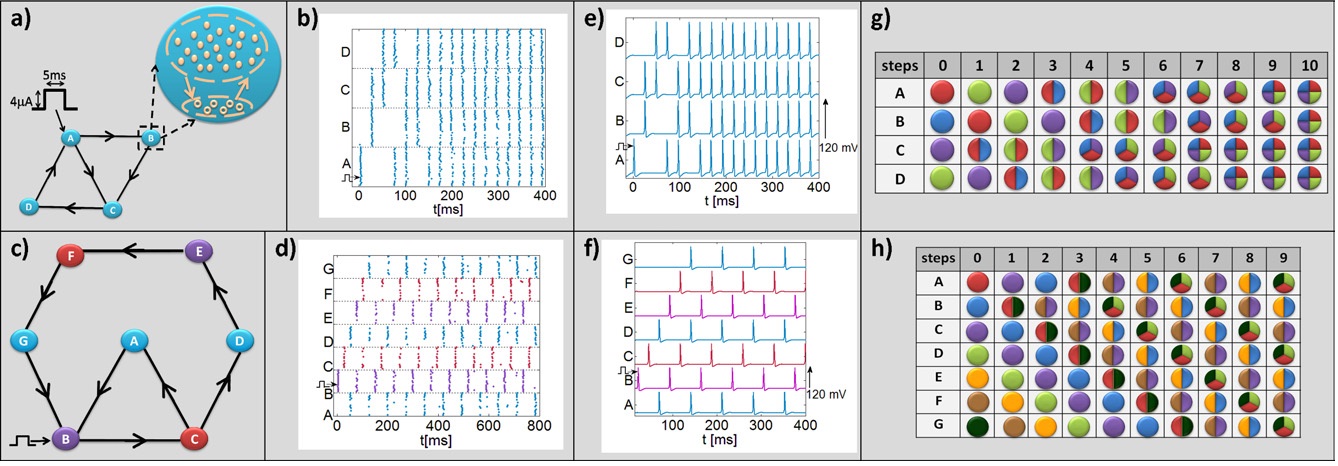}
\end{center}
\caption{ ZLS and clusters in small neural circuits. (a) Schematic of ZLS of an oriented circuit consisting of four nodes, where node A is stimulated for 5 ms by an external current $I_{ext}=4\mu  A/cm^2$. Detailed structure of each node is depicted for node B where filled-circles/empty-circles stand for excitatory/inhibitory neurons and arrows represent reciprocal connections between excitatory and inhibitory neurons. (b) Raster diagram of the firing activity of neurons in (a). (c) Schematic of an oriented circuit consisting of seven nodes which splits into 3-clusters represented by 3 colors. The structure of each node, the distribution of delays and background noise is similar to (a). (d) Raster diagram of the firing activity of neurons in (c). (e) Spike train for (a) with simplified node consisting of only one neuron . (f) Spike train for (c) as for (e). (g) Mixing argument (see text) for (a) where steps are measured in units of $\tau$  and synchronized nodes are composed of an identical set of colors. (h) Mixing argument for (c).
}
\end{figure*}

\section{Neuronal  circuit}
We start with a description of the neuronal circuits and define the properties of a neuronal cell, the structure of a node in a network representing one cortical patch, and the connection between nodes.  Each neural cell was simulated using the well known Hodgkin Huxley model \cite{13} (see Appendix for details). Each node in the network was comprised of a balanced population of 30 neurons, 80/20 percent of which were excitatory/inhibitory (Fig. 1a). The lawful reciprocal connections within each node were only between pairs of excitatory and inhibitory neurons and were selected at random with probability $p_{in}$. In terms of biological properties it was assumed that distant cortico-cortical connections are (almost) exclusively excitatory whereas local connections are both excitatory and inhibitory \cite{15,16}. In this framework, cortical areas are connected reciprocally  across the two hemispheres and within a single hemisphere \cite{16c}, where small functional cortical units (patches) connect to other cortical patches in a pseudo random manner. The number of patches to which a single patch connects  varies considerably, where typically, it grows like the square root of the number of cortical neurons \cite{16a} resulting for a mouse in 3 to 6 \cite{16b}, and most likely for   humans roughly  150. Hence, we investigated excitatory strongly connected oriented graphs; i.e., if neurons belonging to node A project to neurons belonging to node B, then connections from node B to node A are forbidden; however there is a legal path between any pair of nodes \cite{14}. The connection between neurons belonging to different nodes was excitatory and was selected with probability $p_{out}$. In terms of biological properties, distant cortico-cortical unidirectional connections were exclusively excitatory whereas local connections within one node of the network were both excitatory and inhibitory \cite{15,16}.

The delay between a pair of neurons belonging to the same node was taken from a uniform distribution in the range [1.5,2.5] ms, whereas neurons belonging to different nodes came from a uniform distribution in the range $\lbrack \tau-0.5, \tau+0.5 \rbrack$ ms where  $\tau$  was the average time delay including the internal dynamics of a neuron. Results are exemplified below in simulations with  $p_{in}$=0.2, $p_{out}$=0.8 and  $\tau=20$ ms, unless otherwise indicated. The robustness of the results was tested under the influence of background synaptic noise generated from the synaptic input of a balanced random population of 1000 neurons (see Appendix for details). In the absence of a stimulus an isolated node as well as the entire network has no consistent or periodic firing activity or chaotic activity \cite{ido}.

Figure 1a depicts a neuronal circuit consisting of four nodes and two loops having total delays of $3\tau$  and $4\tau$  with GCD(3,4)=1, where at time t=0 node A is stimulated for 5 ms by an external current $I_{ext}=4\mu A/cm^2$. The raster diagram of the firing activity of the neurons in each node is presented in Fig. 1b. Although the graph does not contain reciprocal connections after a short transient,$\sim200$ ms , the neuronal circuits reach ZLS among all nodes.  Figure 1c depicts a neuronal circuit consisting of seven nodes and two loops with total delays of $6\tau$  and $3\tau$  with GCD(6,3)=3, where at time t=0, node B for instance is stimulated for 5 ms by an external current $I_{ext}=4\mu A/cm^2$. The raster diagram of the diluted firing activity of the neurons in each one of the seven nodes is presented in Fig. 1d.  Results indicate that after a short transient of less than 200 ms, the neuronal circuit splits into 3-clusters as labeled by three different colors in Figs. 1c and 1d.

There are two sources for the low firing rate in the neuronal circuits \cite{16b,17a,17b}. The first source is  the local inhibitory connections that hyperpolarized the membrane potential resulting in a relatively diluted firing pattern in each node. To illustrate this effect the Post-Stimulus Time Histogram  (PSTH) of  Fig. 1d is presented in Fig. 2, indicating that at any given time only about half of the population fires simultaneously.  The second source for low firing is the emergence of m-clusters leading to the periodic activity with a period of m$\tau$  of each node rather than $\tau$ as in the case of ZLS. Hence, the cluster mechanism and low firing rate can appear in the brain activity concurrently.  We are well aware that even with the irregularities described here, the statistics of the single neuron firing, as seen here, is far from resembling that of a single cortical neuron in the cortex of a behaving animal.  The model described here is but a toy model compared to the full biological reality.  With more noise, heterogeneous neural properties and more realistic neuronal model (e.g. adding dendrites), much higher irregularities may appear without disrupting the main results presented here: The behavior of a network of connected nodes is an emergent property of all the inter-nodal connections.  Thus, the suitability to describe irregular firing patterns of cerebral cortical area \cite{17a,17b} needs a further investigation and might requires averaging out the regular firing by looking at long recording times.

\begin{figure}[t]
\begin{center}
\includegraphics[angle=0,width=0.32\textwidth,totalheight=0.32\textwidth]{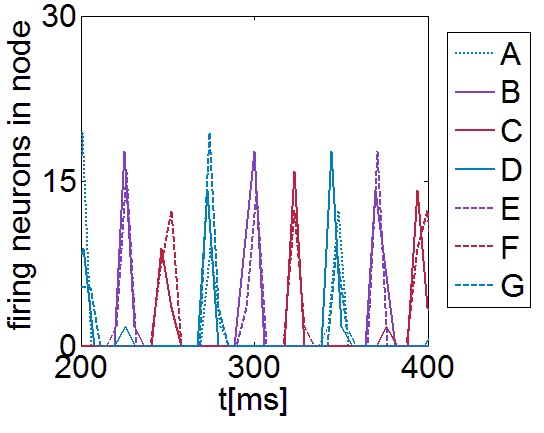}
\end{center}
\caption{ PSTH of the spike trains corresponding to Fig. 1b. The emergence of clustering is demonstrated by the synchronized firing for nodes belonging to a cluster ((C,F), (A,D,G) and (B,E)). The synaptic noise in the circuit, as well as the inhibitory connections within each node, result in low firing dynamics, demonstrated by the fact that at any given time only about half of the population of each node (30 neurons) fires simultaneously.}

\end{figure}

\section{GCD-clusters}
In the absence of inhibition and background noise, the firing pattern of each neuron becomes regular and all neurons in a node are synchronized; however, the phase among nodes is unaltered. For the sake of clarity, below we present results for neural circuits where each node was reduced to one excitatory neuron with no background noise, although similar results were obtained in simulations for structured nodes and background noise.
The results of this simplified node characterized by a single neuron for the neural circuits in Figs. 1a and 1c are presented in Figs. 1e (ZLS) and 1f (3-clusters), respectively.
The interplay between the number of clusters and the GCD of the loops that compose a neuronal circuit  can best be understood by the self-consistent argument that nodes with identical color are in ZLS and must be driven by the same set of "colors". The trivial solution is always one "color", ZLS; however, the alternative solution consists of exactly GCD "colors", GCD-clusters. An attempt to consistently color nodes serially with a greater number of colors fails, because nodes of the same color have different drives. In the case of  $GCD > 1$, GCD-clusters take over the ZLS solution following the mixing argument \cite{18}; the initial condition is a distinct color for each node, time steps are rescaled with $\tau$ and at each time step a node is colored by the union of colors of its driven nodes. The colors of a node at step t indicate the set of nodes/colors at t=0 which are now mixed (integrated) by the node. The mixing argument for the circuit in Fig. 1a is shown in Fig. 1g where after 10 steps all nodes are identical, and colored by 4 colors, indicating a ZLS solution. Similarly the mixing argument for the circuit in Fig. 1c indicates 3-clusters ((A,D,G), (B,E), (C,F)) as depicted in Fig. 1h.
Note that it is possible to consistently color nodes with any common divisor; however, such dynamics requires complex stimuli to more than one node, identical color for nodes in the the initial condition.

\section{Nonlocal mechanism}
A more complex circuit is presented in Fig. 3a consisting of three directed loops with total delays of $6\tau$ , $12\tau$  and $18\tau$ and 25 nodes. The GCD(6,12,18)=6 and 6-clusters (6-colors) were identified in simulations. Small changes in topology can dramatically alter the number of clusters, such that the addition/deletion of one connection can serve as a remote switching mechanism in the circuit \cite{19,20}. Figures 3b, 3c and 2d show that an additional unidirectional connection between two nodes induces a loop with a total delay of $5\tau$ , $4\tau$  and $3\tau$ , respectively. Hence the GCD modifies and switches the 6-cluster solutions to ZLS, 2-clusters and 3-clusters, respectively.  One might conclude that in order to generate large loops the number of cortical patches, nodes, needs to increase accordingly, and furthermore that a circuit composed of loops which are only multiples of a given integer (e.g.  6, 12, 18 $.~.~.$) might be far removed from cortical anatomy.  Figure 3e indicates that an oriented graph consisting of 6 nodes with diluted connectivity generates loops of sizes $6n\tau$ where n is an integer. Furthermore a shortcut in this condensed representation, as shown by the dashed arrow in Fig. 3e, changes the GCD in a way similar to the expanded representation, Figs 3a-d.  We note that a straight chain of neuronal groups with feed forward connections acts like a synfire chain and appropriate connectivity may be found abundantly in the cortex \cite{15}, however the behavior of circuits was not studied so far.

\begin{figure}[h]
\begin{center}
\includegraphics[width=0.6\textwidth,height=0.9\textwidth]{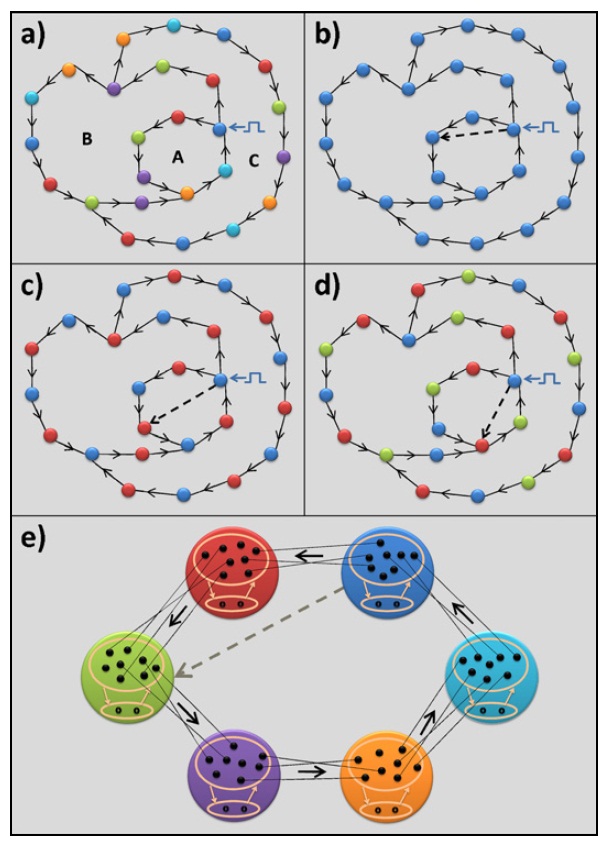}
\end{center}
\caption{ Non-local mechanism for clustering of complex and condensed circuits. (a) A circuit consisting of 25 simplified nodes, loops with total delays of $6\tau$ , $12\tau$  and $18\tau$  (boundaries of areas A, A+B and C, respectively where  $\tau$ is a unit delay between two connected nodes), and a stimulus to one node as in Fig. 1. Nodes split into 6-clusters following the GCD(6,12,18)=6. (b) With an additional unidirectional connection (dashed arrow) and a loop of $5\tau$ , the GCD(5,6,12,18)=1 and the circuit is in ZLS. (c) An additional loop of $4\tau$  as in (b), where GCD(4,6,12,18)=2 and the circuit is in 2 clusters. (d) An additional loop of $3\tau$  as in (b), where GCD(3,6,12,18)=3 and the circuit is in 3 clusters. (e) Schematic condensed representation of 6n  loops where n is an integer, emerging in a ring of 6 nodes, each one representing one cortical patch, and with diluted connectivity where the in/out connectivity of each neuron is at least 1. The gray dashed arrow depicts a similar effect as in 2b.}
\end{figure}

\section{Complex external stimuli}
A temporal stimulus to one node of the neural circuit splits the firing pattern of nodes into GCD-clusters and in addition the firing pattern cycle \cite{22} of a node is also equal to the GCD. For a neural circuit with GCD=6 (Figs. 3a and 3e), the firing pattern cycle of a node is exemplified in the first row of Fig. 4a together with its binary representation and the degeneracy of this class of stimuli.  Under cyclic permutation symmetry in the 6-clusters arrangement, there are 13 classes of simultaneous stimuli for nodes belonging to one or more clusters as summarized by the first column of Fig. 4a, and as expected, the sum of their degeneracy is $2^6-1=63$. The firing pattern cycle and its binary representation are given by the second and third columns of Fig. 4a.

\begin{figure}[h]
\begin{center}
\includegraphics[width=0.60\textwidth,height=0.52275\textwidth]{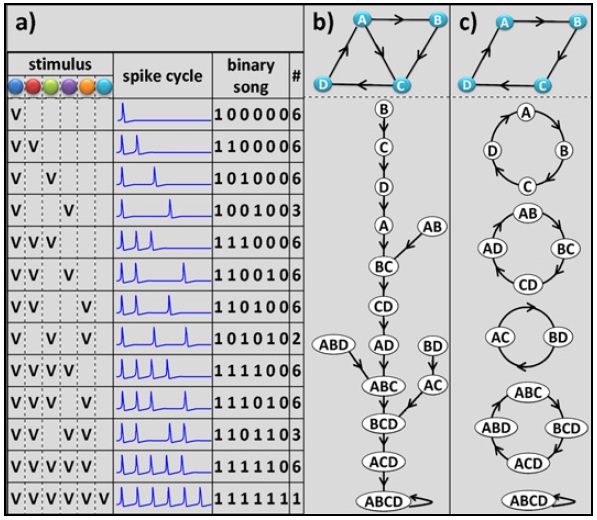}
\end{center}
\caption{ Complex external stimuli and transients. (a) Under permutation symmetry, the 63 different stimuli in the 6-cluster arrangement of the circuit in Fig. 3a are organized into 13 classes (first column), where the degeneracy of each class is given in the last column. The firing pattern cycle of a node together with its binary representation are presented in the second and the third columns, indicating that the number of clusters as well as the firing spike cycle of a node can be any common divisor of the loops. (b) Four node circuit which is in ZLS and the transients for the 15 different stimuli organized in a tree. (c) A square circuit and its 15 different stimuli organized in 5 distinct cyclic flows.
}
\end{figure}

Results indicate that the period of the firing pattern can differ from the GCD (found also for chaotic networks \cite{ido}), for instance the stimulus period in the $4^{th}$ ( $8^{th}$) row of Fig. 4a is 3 (2) and hence the nodes only split into 3-clusters (2-clusters), instead of 6-clusters. In fact the number of clusters can be any common divisor of the loops composing the circuit as a result of stimuli inducing such periodicity, and an example is presented in Fig. 5.

\begin{figure}[h]
\begin{center}
\includegraphics[width=0.63\textwidth,height=0.549\textwidth]{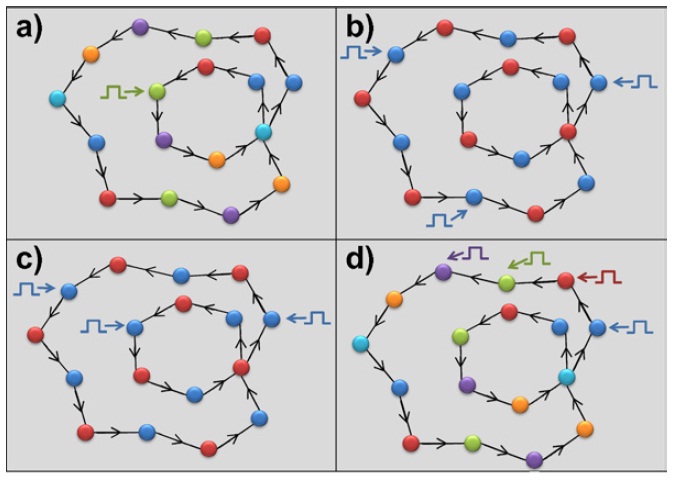}
\end{center}
\caption{ Simulation results of two connected loops with total delays of $6\tau$  and $12\tau$  consisting of 17 nodes. (a) A drive to one node resulting in GCD(6,12)=6 clusters.  (b) A drive to every fourth node in the loop of 12 , represented by $(1,0,0,0,1,0,0,0,1,0,0,0)$, results in GCD(4,6,12)=2 clusters. (c) The drive can be given to any set of nodes with the same colors as in (b) (blue, orange, green) and results in the same clusters. (d) A drive to four consecutive nodes in the loop of 12  represented by $(1,1,1,1,0,0,0,0,0,0,0,0)$, which is characterized by a periodicity of 12. Hence, it results in GCD(6,12)=6 clusters.
}
\end{figure}

The number of clusters as well as the spiking cycle of a given node can identify a class of possible stimuli applied to the circuits. Nevertheless, more detailed information about the stimuli can be deduced rapidly from the transients \cite{24} to synchronization. Fig. 4b shows all the transients to ZLS from the 15 possible stimuli for the circuit in Fig. 1a, whereas Fig. 4c presents the 5 possible cyclic flows for a unidirectional square circuit. It is clear that the length and even knowledge of a partial time ordering of firing nodes in the transient quite clearly identify the stimulus, and the synchronized mode mainly serves as an indicator of the end of the transient.

\begin{figure}[h!]
\begin{center}
\includegraphics[width=0.6\textwidth,height=0.36\textwidth]{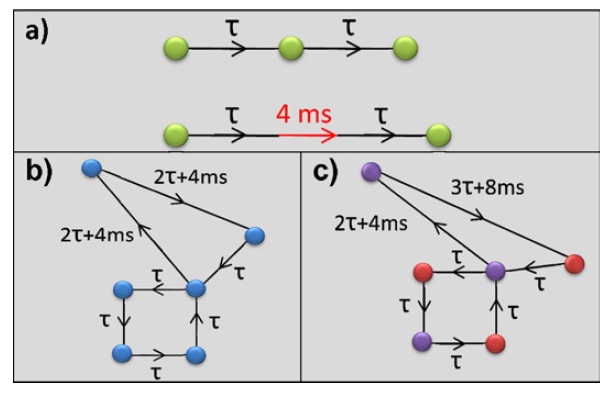}
\end{center}
\caption{ ZLS and clusters in heterogeneous circuits. (a) A chain of three neurons with an accumulated $2\tau$  ms delay is equivalent to a chain of two neurons with a $2\tau +4$ ms delay as a result of the 4 ms internal dynamic of the middle neuron.  (b) Heterogeneous circuit at ZLS which is equivalent to a homogeneous circuit consisting of $4\tau$  and $5\tau$  connecting loops. (c) Heterogeneous circuit which is equivalent to a homogeneous circuit consisting of $4\tau$  and $6\tau$  connecting loops, 9 nodes and GCD(4,6)=2. The 6 nodes of the heterogeneous circuit are colored according the corresponding nodes in the homogeneous circuit.}
\end{figure}

\section{Heterogenous circuits}
Generalization of the above homogeneous circuit results to heterogeneous circuits is depicted in Fig. 6. Figure 5a indicates that the internal dynamic of a neuron results in an effective delay of about 4 ms. Hence, a chain of three neurons with a total delay of $2\tau$  is equivalent to a chain of two units with one delay of $2\tau +4$ ms and vice versa. Consequently, Fig. 6b presents a heterogeneous circuit which is equivalent to a homogeneous circuit composed of two loops of $4\tau$  and $5\tau$ . Since GCD(4,5)=1, the homogeneous as well as the heterogeneous circuits are in ZLS. Similarly, the delay of $2\tau +4$ ($3\tau +8$) in Fig. 6c can be replaced by  homogeneous chains of 3 (4) nodes and the circuit is equivalent to two homogeneous loops of $4\tau$  and $6\tau$. Since GCD(4,6)=2, the original  6 nodes in the heterogeneous circuit, Fig. 6c, are colored according to the corresponding nodes in the homogeneous circuit.

The 4 ms internal dynamic of a neuron was estimated in simulations for large loops. However, for shorter loops, e.g., 6 units, the mapping between homogeneous/heterogeneous networks is found to valid in simulations when the internal dynamics is assumed to be in the range of $\lbrack 2,5 \rbrack$ ms.  This range enables a robust synchronization mode of heterogeneous circuits with a wide range of delays, as well as heterogeneous circuits with complex nodes and fluctuating delays as in Fig. 1a.

\section{Concluding remarks}
The activity mode of the entire network cannot simply be described as a "Lego" of small connecting neural circuits with a given activity, since it is governed by a nonlocal quantity, the GCD. These findings challenge the emergence of significant topological motifs and the importance of their role in the functionality of the entire network \cite{26} as well as the impact of statistical properties of complex neural circuits \cite{27}. Rather, they call for a reexamination of sources of correlated activity in cortex where addition/deletion of a connection or more realistically synaptic alternations that can induce transition between decaying and sustained activities can serve as a remote switching mechanism, and indicate that learning induced changes in some connections affects the functionality of the networks more than others. The hypothesis that neural information processing might take place in the transient is suggestive of a much shorter time scale for the inference of a perceptual entity, which might also indicate the emergence of a probabilistic inference consecutive to the accumulation of knowledge during the transient. The notion that complex external stimuli result in several modes of activity also implies richer capabilities of the neural circuit and makes it imperative to analyze complex stimuli that are not triggered simultaneously. Nevertheless, the ways in which neural circuits detect synchronization still remain an enigma.

\section{Appendix}
To simulate neural population dynamics we used  Hodgkin-Huxley (HH) type models for action potential generation, in which the membrane potential $V^i$ , of a single neuron $i$ , is described by the following differential equation
\begin{align}
c_m {dV^i \over dt}= & -g_{Na}(m^i)^3h^i(V^i-E_{Na}) \nonumber \\ & -g_k(n^i)^4(V^i-E_k) \nonumber \\ & -g_L(V^i-E_L)+I^i_{syn}+I^i_{ext}+I^i_{noise}
\end{align}
\noindent Here, $c_m=1\mu F/cm^2$ is the membrane capacitance. Maximal conductance and reversal potentials are given by:

\begin{table} [h!]
\begin{center}
\begin{tabular} {|c | c | c |}
\hline
$X  $ & $  g_{X} [mS/cm^{2}]$ & $  E_{X} [mV] $ \\ \hline
$N_{a}$ & $120$ & $115$ \\ \hline
$k$ & $36$ & $-12$ \\ \hline
$L$ & $0.3$ & $10.5$ \\ \hline
\end{tabular}
\end{center}
\end{table}

The voltage-gated ion channels $m$, $n$ represent the activation of the sodium and potassium channels and the voltage-gated ion channel, $h$, represent inactivation of the sodium channels. Their dynamics is described by first order kinetic equations for a given neuron $i$:
\begin{equation}
{dY^i \over dt}=\alpha^i_Y(V^i)(1-Y^i)-\beta^i_Y(V^i)Y^i
\end{equation}
\noindent where Y may be substituted by $m$, $n$ and $h$. The experimentally fitted voltage-dependent transition rates are:

\begin{eqnarray}
&\alpha_{m}(V)= \frac{0.1 (V-25)}{1-e^{-0.1(V-25)}} \nonumber\\
&\beta_{m}(V)= 4 e^{-V/18} \nonumber\\
&\alpha_{n}(V)= \frac{0.1 (V-10)}{1-e^{-0.1(V-10)}} \nonumber\\
&\beta_{n}(V)= 0.125 e^{-V/80} \nonumber\\
&\alpha_{h}(V)= 0.07 e^{-V/20} \nonumber\\
&\beta_{h}(V)= \frac{1}{1+e^{-0.1(V-30)}} \nonumber
\end{eqnarray}

Synaptic background noise was simulated by a balance of input from 800 excitatory neurons firing at about 1-3Hz, and 200 inhibitory neurons firing at about 50-100 Hz. Our simulations were adjusted to imitate the behavior of a random biological neural cell. The free parameters were set in a way such that a single cell with no synaptic or external input fired randomly at about 5-7 Hz and the cell activity was noisy around the resting potential with a variance of about 5 mV.
The synaptic transmission between neurons was modeled by a postsynaptic conductance change in the form of an $\alpha$ function

\begin{equation}
\alpha (t) = \frac{e^{-t / \tau_{d}} - e^{-t / \tau_{r}}}{\tau_d - \tau_r}
\end{equation}

\noindent where the decay and rise time of the function are given by  $\tau_d=10$ ms and  $\tau_r=1$ ms  respectively. The synaptic current $I^{i}_{syn}(t)$ takes the form
\begin{equation}
I^{i}_{syn}(t) = -g_{max} \sum_{j} \sum_{t_{j}^{sp}} \alpha (t- t_{j}^{sp} -\tau_{ij}) (V-E_{syn})
\end{equation}

\noindent Here, $\{j\}$ is the group of neurons coupled to neuron i. The internal sum is taken over the train of pre-synaptic spikes occurring at $t^{sp}_{j}$  of a neuron j  in the group. The delays arising from the finite conduction velocity of axons are taken into account through the latency time  $\tau_{ij}$  in the $\alpha$ function. Excitatory and inhibitory transmissions were differentiated by setting the synaptic reversal potential to be $E_{syn}=60$ mV or $E_{syn}=-20$ mV , respectively. $g_{max}$ describes the maximal synaptic conductance between neurons i and j. We integrated the set of differential equations numerically using Heun's method. The time step of the integration was 0.02ms.

\textbf{Population dynamics:}
Each node in the circuit was considered as one cortical area and comprised a balanced population of 30 excitatory and inhibitory neurons that were coupled sparsely with a fixed probability $p_{in}=0.8$ for every connection between two neurons. Every two neurons that belonged to adjacent nodes were connected with a fixed probability $p_{out}$ and connection strength $g_{max}=0.17 mS/cm^2$. The delays between groups of neurons were taken as a distribution around a mean value. Neurons belonging to the same population had an average delay time of 2 ms and neurons from different populations had delays of 20 ms.

\end{document}